\documentclass[12pt]{article}
\usepackage[dvips]{graphicx}
\usepackage{amssymb}

\makeatletter
 \@addtoreset{equation}{section}
 
\makeatother

\newcommand{\maru}{\mbox{\tiny$\stackrel{\circ}{\scriptstyle\circ}$}}

\begin{document}


\begin{titlepage}

\renewcommand{\thefootnote}{\fnsymbol{footnote}}

\begin{flushright}
\begin{tabular}{l}
UTHEP-541\\
KEK-TH-1142\\
hep-th/0703216 \\
March 2007
\end{tabular}
\end{flushright}

\bigskip

\begin{center}
{\Large \bf Observables and Correlation Functions
  in $OSp$ Invariant String Field Theory}\\
\end{center}

\bigskip

\begin{center}
{\large
Yutaka Baba}${}^{a}$\footnote{e-mail:
        yutaka@het.ph.tsukuba.ac.jp},
{\large Nobuyuki Ishibashi}${}^{a}$\footnote{e-mail:
        ishibash@het.ph.tsukuba.ac.jp},
{\large Koichi Murakami}${}^{b}$\footnote{e-mail:
        koichi@post.kek.jp}
\end{center}

\begin{center}
${}^{a}${\it
Institute of Physics, University of Tsukuba,\\
Tsukuba, Ibaraki 305-8571, Japan}\\
\end{center}
\begin{center}
${}^{b}${\it High Energy Accelerator Research Organization (KEK),\\
Tsukuba, Ibaraki 305-0801, Japan}
\end{center}

\bigskip

\bigskip

\bigskip

\begin{abstract}
We define BRST invariant observables in the $OSp$ invariant
closed string field theory for bosonic strings.
We evaluate correlation functions of these observables
and show that the S-matrix elements derived from them
coincide with those of the light-cone gauge string field
theory.
\end{abstract}

\setcounter{footnote}{0}
\renewcommand{\thefootnote}{\arabic{footnote}}

\end{titlepage}

\section{Introduction}

The $OSp$ invariant string field theory 
\cite{Siegel:1984ap}\cite{Neveu:1986cu}\cite{Kugo:1987rq}
is a covariantized version 
of the light-cone gauge string field theory 
\cite{Kaku:1974zz}\cite{Mandelstam:1973jk}\cite{Cremmer:1974ej}.
It is made to reproduce the results of light-cone gauge string field 
theory via Parisi-Sourlas mechanism \cite{Parisi:1979ka}. 
Involving extra time and length variables, 
the formulation of the theory is not similar to the usual ones 
but rather like stochastic quantization. 
For noncritical strings, stochastic type formulation of 
string field theory was proposed to reproduce the results 
of matrix models \cite{Ishibashi:1993pc}. 
Therefore it is likely that what can be done for noncritical 
strings can be done also for critical strings by using this 
$OSp$ invariant string field theory formulation. 
Indeed, 
using the results of noncritical string theories
\cite{Fukuma:1996hj}\cite{Hanada:2004im}
and idempotency of boundary states 
\cite{Kishimoto:2003ru}\cite{Kishimoto:2004jk},
we constructed solitonic operators
which can be regarded as D-branes in the 
$OSp$ invariant string field theory in \cite{Baba:2006rs}.

Therefore the $OSp$ invariant string field theory may be useful 
to study the nonperturbative effects involving D-branes. 
However, since the structure of the action is quite different 
from that of usual string field theories, it is not easy to 
see how the closed string particle modes are realized in 
this string field theory. 
What we would like to do in this paper is to clarify this point. 
We will consider the $OSp$ invariant string field theory for closed bosonic 
strings and  define BRST invariant observables corresponding to these 
particle modes and study their correlation functions. 
We will show that the S-matrix elements can be derived 
from the correlation functions and 
they coincide with those of the light-cone gauge string field theory. 

Our treatment is different from the previous ones 
\cite{Neveu:1986cu}\cite{Kugo:1987rq} in which 
on-shell physical states are considered. 
Since the kinetic term of the action is not similar to that of the usual 
formulation, it is difficult to fix the normalization
of these states.\footnote{In \cite{Kugo:1987rq},
this problem was addressed, and solved for covariantized 
light-cone string field theory \cite{Uehara:1987qz}\cite{Kawano:1992dp}.} 
By considering the observables instead, we can fix the normalization 
using the two-point correlation functions. 
Another advantage of our method is
that we can show the Parisi-Sourlas reduction 
without Euclideanizing the $\pm$ directions. 
The $OSp$ invariant string field theory is inherently Lorentzian
in these directions 
and the BRST cohomology is defined for such signature. 
Therefore it is important to show that the reduction occurs
without changing the signature. 


The organization of this paper is as follows.
In section \ref{sec:OSp-inv-SFT},
we will review the $OSp$ invariant string field theory.
In section \ref{sec:cohomology}, we will define 
the observables of the $OSp$ invariant string field theory.
In section \ref{sec:S-matrix-equivalence},
we will study the correlation functions of the observables 
defined in section \ref{sec:cohomology}, 
and show that the S-matrix elements which can be derived 
from these correlation functions 
coincide with those in the light-cone gauge string field theory.
Section \ref{sec:discussions} will be devoted to discussions.

In this paper, we set the string slope parameter
$\alpha'$ to be 2.

\section{$OSp$ Invariant String Field Theory}
\label{sec:OSp-inv-SFT}

In order to fix the notations,
we review the $OSp$ invariant string field theory.\footnote{
In this paper, we use conventions slightly different from
those of our previous paper \cite{Baba:2006rs},
in particular that for the integration measure
of the momentum zero-modes.
}

The procedure of \cite{Siegel:1984ap} for covariantizing
the light-cone gauge string field
theory \cite{Kaku:1974zz}\cite{Mandelstam:1973jk}%
\cite{Cremmer:1974ej} 
is to replace the $O(24)$ transverse vector $X^{i}$ $(i=1,\ldots, 24)$
by the $OSp(26|2)$ vector
$X^{M}=\left( X^\mu ,C,\bar{C}\right)$,
where $X^{\mu}=(X^{i},X^{25},X^{26})$ are Grassmann even
and the ghost fields $C$ and $\bar{C}$ are Grassmann odd.
The metric of the $OSp(26|2)$ vector space
is\footnote{In this paper, we begin by the Euclidean signature
for the metric in the linearly realized $O(26)$ directions.
This is different from the original formulation in \cite{Siegel:1984ap}
where the signature of the metric in these directions is Lorentzian.
}
\begin{equation}
\eta_{MN}
=
\begin{array}{c}
\\
\\
\\
\\
\mbox{\scriptsize $C$}\\
\mbox{\scriptsize $\bar{C}$}\\
\end{array}
\begin{array}{r}
\mbox{\scriptsize$C$}~~\mbox{\scriptsize$\bar{C}$}\hspace{6mm}\\
\left(
\begin{array}{ccc|cc}
 &             & & &  \\
 &\delta_{\mu\nu}& & &  \\
 &             & & &  \\\hline
 &             & &0&-i\\
 &             & &i&0 \\
\end{array}
\right)
\end{array}
 =\eta^{MN}~.
\end{equation}
In accordance with the above $OSp$ extension, we extend
the oscillation modes $\alpha_{n}^{i}$
and $\tilde{\alpha}^{i}_{n}$
$(i=1,\ldots,24; n\in \mathbb{Z})$
as well in the following way,
\begin{eqnarray}
 &&
   x^{i} \longrightarrow x^{M}
   =\left(x^{\mu},C_{0},\bar{C}_{0}\right)~,
  \nonumber\\
 &&
  \alpha^{i}_{0} = \tilde{\alpha}^{i}_{0}
  = p^{i}
  \longrightarrow
  \alpha^{M}_{0}=\tilde{\alpha}^{M}_{0}
  =p^M
  =\left(p^{\mu},-\pi_{0},\bar{\pi}_{0}
   \right)~,
  \nonumber\\
 &&
\alpha^{i}_{n} \longrightarrow
\alpha^{M}_{n}
  =\left(\alpha^{\mu}_{n},-\gamma_{n},\bar{\gamma}_{n}\right)~,
 \quad
\tilde{\alpha}^{i}_{n} \longrightarrow
   \tilde{\alpha}^{M}_{n}
   =\left(\tilde{\alpha}^{\mu}_{n},-\tilde{\gamma}_{n},
          \tilde{\bar{\gamma}}_{n}\right)
 \quad \mbox{for $n\neq 0$}~.
\end{eqnarray}
These oscillators satisfy the canonical commutation relations
\begin{equation}
[x^{N},p^{M}\}=i\eta^{NM}~,
\quad [\alpha^{N}_{n},\alpha^{M}_{m} \}=n\eta^{NM}\delta_{n+m,0}~,
\quad [\tilde{\alpha}^{N}_{n},\tilde{\alpha}^{M}_{m} \}
  = n \eta^{NM} \delta_{n+m,0}
\end{equation}
for $n,m\neq 0$,
where the graded commutator $[A,B\}$ denotes the anti-commutator
when $A$ and $B$ are both fermionic operators and
the commutator otherwise.

We describe the Hilbert space for the string by the
Fock space of the oscillators for the non-zero modes
and the wave functions for the zero-modes.
We take the momentum representation of the wave functions 
for the zero-modes
$p^\mu , \alpha , \pi_0, \bar{\pi}_0$,
where $\alpha$ is identified with the string length.
In this description,
the vacuum state $|0\rangle$ in the first quantization
is defined by
\begin{equation}
x^{M}|0\rangle
 =i\eta^{MN}\frac{\partial}{\partial p^{N}} |0\rangle
 =0~,
\quad
\alpha^{M}_{n}|0\rangle=\tilde{\alpha}^{M}_{n}|0\rangle
 = 0 \qquad \mbox{for $n>0$}~.
 \label{eq:vacuum1stq}
\end{equation}
The integration measure for the zero-modes 
of the $r$-th string is defined as 
\begin{equation}
dr
\equiv
 \frac{\alpha_rd\alpha_r}{2}
 \frac{d^{26}p_r}{(2\pi)^{26}} \, 
i  d\bar{\pi}_0^{(r)} d\pi_0^{(r)}~.
 \label{eq:zeromodemeasure}
\end{equation}

The action of the $OSp$ invariant string field theory
is obtained by the $OSp$ extension explained above
from that of the light-cone gauge string field theory
given in \cite{Kaku:1974zz}.
This takes the form
\begin{eqnarray}
\lefteqn{
S = \int dt  \left[
  \frac{1}{2}
   \int d1d2\, \left\langle R(1,2) \left|\Phi\right\rangle_{1}
               \right.
       \left( i\frac{\partial}{\partial t}
                  -\frac{L_0^{(2)}+\tilde{L}_0^{(2)}-2}{\alpha_2}
       \right)  \left|\Phi\right\rangle_{2}
   \right.} \hspace{3.5em}\nonumber\\
&& \left.
{} + \frac{2g}{3}
\int d1d2d3 \, \left\langle V_3^0(1,2,3)\right|
\Phi\rangle_1|\Phi\rangle_2|\Phi\rangle_3
\right]~,
\label{eq:actionOSp}
\end{eqnarray}
where 
$\langle R(1,2)|$ is the reflector defined in eq.(\ref{eq:reflector})
and $\langle V_{3}^0(1,2,3)|$ is the three-string vertex
given as
\begin{equation}
\left\langle V_{3}^{0}(1,2,3) \right|
 \equiv  \delta (1,2,3)
   \; {}_{123}\!\langle 0|e^{E (1,2,3)}
    \mathcal{P}_{123}
    \frac{|\mu (1,2,3)|^2}{\alpha_1\alpha_2\alpha_3}~.
\label{V30}
\end{equation}
$E(1,2,3)$, $\mathcal{P}_{123}$, $\delta(1,2,3)$
and $\mu(1,2,3)$ in this equation
are defined in eq.(\ref{eq:V3variables}).
The string field $\Phi$ is taken to be Grassmann even
and subject to the level matching condition
$\mathcal{P}\Phi=\Phi$ and the
reality condition
\begin{equation}
\langle \Phi_{\mathrm{hc}} |
  = \langle \Phi |~,
  \label{eq:reality1}
\end{equation}
where
$\langle \Phi_{\mathrm{hc}} |
\equiv \left( | \Phi \rangle \right)^{\dagger}$
denotes the hermitian conjugate of
$| \Phi \rangle$,
and $\langle \Phi |$ denotes the BPZ conjugate
of $|\Phi \rangle$ defined in eq.(\ref{eq:BPZ}).

The action (\ref{eq:actionOSp}) is
invariant under the BRST transformation
\begin{equation}
\delta_{\mathrm{B}} \Phi
 =Q_{\mathrm{B}} \Phi +g\Phi *\Phi~,
\label{BRST}
\end{equation}
where the $\ast$-product is given in eq.(\ref{eq:star-prod})
and the BRST operator
$Q_{\mathrm{B}}$ is defined
\cite{Siegel:1986zi}\cite{Bengtsson:1986yj} as 
\begin{eqnarray}
 Q_{\mathrm{B}}
 &=& \frac{C_0}{2\alpha}(L_0+\tilde{L}_{0} - 2)
   -i\pi_0\frac{\partial}{\partial \alpha}
  \nonumber\\
&& 
    {} +\frac{i}{\alpha}\sum_{n=1}^{\infty}
      \left(
        \frac{\gamma_{-n}L_n-L_{-n}\gamma_n}{n}
          +\frac{\tilde{\gamma}_{-n}\tilde{L}_n
                 -\tilde{L}_{-n}\tilde{\gamma}_n}{n}\right)~.
  \label{eq:brst-charge}
\end{eqnarray}
Here $L_{n}$ and $\tilde{L}_{n}$ $(n\in\mathbb{Z})$ are
the Virasoro generators given by
\begin{equation}
L_n = \frac{1}{2} \sum_{m\in \mathbb{Z}}
     \maru \alpha_{n+m}^{N}\alpha_{-m}^{M} \eta_{NM}\maru~,
\quad
\tilde{L}_n = \frac{1}{2} \sum_{m\in\mathbb{Z}}
      \maru \tilde{\alpha}_{n+m}^{N}
              \tilde{\alpha}_{-m}^{M} \eta_{NM}
         \maru~,
\end{equation}
where the symbol $\maru\ \maru$ denotes
the normal ordering of the oscillators
in which the non-negative modes should be placed to the right
of the negative modes. 
The BRST operator (\ref{eq:brst-charge}) 
can be identified  with the $M^{C-}$ element of 
the $OSp(27,1|2)$ Lorentz generators
\cite{Siegel:1986zi}\cite{Bengtsson:1986yj}.

\section{BRST Cohomology and Observables}
\label{sec:cohomology}

In the $OSp$ invariant string field theory, 
we consider BRST invariant objects as physical quantities. 
The S-matrix elements are defined 
for on-shell BRST invariant states. 
In this section, we will first show that these states correspond to the 
on-shell physical states of string theory. 
Then we define observables from whose correlation functions
we can deduce the S-matrix elements of the $OSp$ invariant theory. 

\subsection{BRST cohomology of $Q_{\mathrm{B}}$}
In order to obtain the BRST cohomology of the asymptotic states
in the $OSp$ invariant string field theory,
we need the BRST cohomology of the operator $Q_{\mathrm{B}}$. 
For studying the BRST cohomology of $Q_{\mathrm{B}}$,
it is convenient to relate $Q_{\mathrm{B}}$ to Kato-Ogawa's
BRST operator~\cite{Kato:1982im}. 
The worldsheet variables in the ghost sector of
the $OSp$ invariant string field theory can be identified 
with the $(b,c)$ ghost variables as 
\begin{eqnarray}
&& C_0 = 2\alpha c_0^+~,
    \quad \bar{\pi}_0 = \frac{1}{2\alpha}b_0^+~,
  \nonumber \\
&& \gamma_n = in\alpha c_n~,
   \quad  \tilde{\gamma}_n = in\alpha\tilde{c}_n~,
   \quad  \bar{\gamma}_n = \frac{1}{\alpha}b_n~,
   \quad \tilde{\bar{\gamma}}_n = \frac{1}{\alpha}\tilde{b}_n
\label{ghostidentification}
\end{eqnarray}
for $n\neq 0$.
{}From this identification, one can find that
the $OSp$ invariant string field theory
includes extra variables
$\pi_0,\alpha$ besides 
those in the usual covariantly quantized theory. 
Here let us introduce $c\equiv \frac{\pi_0}{\alpha}$ for later 
convenience. 
Then the first-quantized Hilbert space of the $OSp$ invariant theory 
is the tensor product of that of the usual covariant string theory
and that of $c,\alpha$. 
The BRST operator can be written as 
\begin{eqnarray}
Q_{\mathrm{B}}=Q_{\mathrm{B}}^{\mathrm{KO}}-ic
   \left(\alpha  \left.\frac{\partial}{\partial\alpha}
              \right|_{b,c}
+1 \right)~,
\label{brst_gamma}
\end{eqnarray}
where $Q_{\mathrm{B}}^{\mathrm{KO}}$ is 
the usual Kato-Ogawa's BRST operator 
with $b_0^-$ omitted
and
$\left.\frac{\partial}{\partial\alpha}
 \right|_{b,c}$
denotes the derivative with
$c,c_0^+,b_0^+,c_n,b_n,\tilde{c}_n,\tilde{b}_n(n\neq 0)$
kept fixed.

Now let us see what a BRST closed state $|~\rangle$ looks like. 
It is convenient to expand $|~\rangle$ 
in $c$:
\begin{equation}
\left|~\right\rangle
=|1\rangle + c |2\rangle~,
\label{eq:state_12}
\end{equation}
where the states $|1\rangle$ and $|2\rangle$ are
independent of $c$.
In this notation,
the condition that 
$|~\rangle $ is $Q_{\mathrm{B}}$-closed
becomes
\begin{eqnarray}
Q_{\mathrm{B}}^{\mathrm{KO}} |1\rangle &=& 0~,
\label{BRST_1}\\
Q_{\mathrm{B}}^{\mathrm{KO}} |2\rangle &=&
\mathcal{D}_\alpha |1\rangle~,
\label{BRST_2}
\end{eqnarray}
where 
\begin{equation}
\mathcal{D}_\alpha
\equiv
-i
\left(\alpha 
     \left.\frac{\partial}{\partial\alpha}\right|_{b,c}
        +1     \right)~.  
\end{equation}

Since we know the BRST cohomology of $Q_{\mathrm{B}}^{\mathrm{KO}}$, 
solutions to eq.(\ref{BRST_1}) can be easily
found to be a linear combination of 
states of the form $f(\alpha )|\mathrm{phys} \rangle $ and 
$g(\alpha )Q_{\mathrm{B}}^{\mathrm{KO}} |~\rangle^\prime $, where 
$|\mathrm{phys} \rangle $ denotes a state
in a nontrivial cohomology class of 
$Q_{\mathrm{B}}^{\mathrm{KO}}$ and $f(\alpha ),g(\alpha )$
are arbitrary functions of $\alpha$. 
Substituting these into eq.(\ref{BRST_2}), one can see
$f(\alpha )=\frac{1}{\alpha}$ and 
$|2\rangle $ should be a linear combination of the solutions to 
\begin{equation}
Q_{\mathrm{B}}^{\mathrm{KO}} |2\rangle =
(\mathcal{D}_\alpha g(\alpha ))
Q_{\mathrm{B}}^{\mathrm{KO}} |~\rangle^\prime~.
\end{equation}
Solutions to this equation can also be easily found
and eventually we see that the BRST closed state $|~\rangle$
should be a linear combination of the states of the form 
\begin{equation}
\frac{1}{\alpha}|\mathrm{phys}\rangle~,
\label{brsclosed1}
\end{equation}
and
\begin{equation}
c h(\alpha )|\mathrm{phys}\rangle~,
\label{brsclosed2}
\end{equation}
up to $Q_{\mathrm{B}}$ exact states. 
Here $h(\alpha )$ is an arbitrary function of $\alpha$. 

For $|\mathrm{phys}\rangle$, one can choose the states of the form 
\begin{equation}
|0\rangle_{b,c}\otimes 
|\overline{\mathrm{primary}}\rangle_X
(2\pi )^{26}\delta (p-k)
~,
\end{equation}
or
\begin{equation}
b_0^+|0\rangle_{b,c}\otimes 
|\overline{\mathrm{primary}}\rangle_X
(2\pi )^{26}\delta (p-k)
~,
\end{equation}
where $|0\rangle_{b,c}$ is a vacuum for the $(b,c)$ ghosts satisfying 
$c_0^+|0\rangle_{b,c}=0$.\footnote{Notice that $b_0^-,c_0^-$ are omitted.} 
$|\overline{\mathrm{primary}}\rangle_X$ is the oscillator part of a 
Virasoro primary state in the Hilbert space of $X^{\mu}$ variables 
and $k^\mu$ is the momentum eigenvalue. 
In order for these states to be BRST closed and nontrivial, 
the conformal weight of the 
the primary state $|\overline{\mathrm{primary}}\rangle_X(2\pi )^{26}(p-k)$ 
should be $(1,1)$. 
This condition can be regarded as the on-shell condition
for the particle corresponding to this string state.

\subsubsection*{$\alpha$-dependence}
The wave functions for the $OSp$ invariant string field theory
should satisfy appropriate boundary conditions. 
Especially one should be careful about the dependence
on the zero-mode $\alpha$. 
Treating the regions $\alpha >0$ and $\alpha <0$ separately,
let us introduce a real 
variable $\omega$ as
\begin{equation}
\alpha =\pm e^{\omega}.
\end{equation}
If we express the wave functions using the original variables
in the $OSp$ invariant string field theory, 
the $\alpha$ dependent part of the wave functions should be of the form
\begin{equation}
e^{-\omega}f(\omega )~,
\label{wavealpha}
\end{equation}
where $f(\omega )$ is a delta function normalizable function 
with respect to the norm
\begin{equation}
\| f \|^2
\equiv
\int_{-\infty}^\infty d\omega |f(\omega )|^2~.
\end{equation}
We can take $e^{i\omega x}$ $(x \in \mathbb{R})$ as 
a basis for such wave functions. 
It is straightforward to show that under such conditions 
$Q_{\mathrm{B}}$ and $M^{+-}$ are hermitian. 
If we express the wave functions using the $(b,c)$ ghosts,
$\alpha$ and $c$, 
eq.(\ref{wavealpha}) should be replaced by 
\begin{equation}
e^{(n-1)\omega}f(\omega )~,
\end{equation}
where $n$ is the ghost number of the state. 
The ghost number is defined so that the variable $c$ has ghost number $1$ 
and the state $|0\rangle_{b,c}$ has ghost number $0$.

Now let us take this condition into account and further restrict the form 
of the BRST closed states. 
For the states of the form in eq.(\ref{brsclosed1}), 
$|\mathrm{phys}\rangle$ should have ghost number $0$ and 
therefore it should be of the form
\begin{equation}
\frac{1}{\alpha}|0\rangle_{b,c}\otimes 
|\overline{\mathrm{primary}}\rangle_X
(2\pi )^{26}\delta (p-k)
~.
\end{equation}
The states of the form in eq.(\ref{brsclosed2}) should be either 
\begin{equation}
ce^{i\omega x}|0\rangle_{b,c}\otimes 
|\overline{\mathrm{primary}}\rangle_X
(2\pi )^{26}\delta (p-k)
~,
\end{equation}
or
\begin{equation}
\frac{b_0^+c}{\alpha}e^{i\omega x}|0\rangle_{b,c}\otimes 
|\overline{\mathrm{primary}}\rangle_X
(2\pi )^{26}\delta (p-k)
~,
\end{equation}
but the former one is BRST exact and the latter is BRST exact if $x\neq 0$. 
Therefore we have shown that the BRST closed states can be written 
as a linear combination of the states of the form 
\begin{equation}
\frac{1}{\alpha}|0\rangle_{b,c}\otimes |\overline{\mathrm{primary}}\rangle_X
(2\pi )^{26}\delta (p-k)
~,
\end{equation}
and
\begin{equation}
\frac{b_0^+c}{\alpha}|0\rangle_{b,c} \otimes 
  |\overline{\mathrm{primary}}\rangle_X
  (2\pi )^{26}\delta (p-k)
~,
\end{equation}
up to BRST exact states. 
Written in terms of the original variables of the $OSp$ theory, these are
\begin{equation}
\frac{1}{\alpha}|0\rangle_{C,\bar{C}}\otimes 
|\overline{\mathrm{primary}}\rangle_X
(2\pi )^{26}\delta (p-k)
~,
\label{brscohomology1}
\end{equation}
and
\begin{equation}
\frac{1}{\alpha}\bar{\pi}_0\pi_0|0\rangle_{C,\bar{C}}\otimes 
|\overline{\mathrm{primary}}\rangle_X
(2\pi )^{26}\delta (p-k)
~,
\label{brscohomology2}
\end{equation}
where 
$|0\rangle_{C,\bar{C}}$ is the oscillator 
vacuum (\ref{eq:vacuum1stq}) for the $C,\bar{C}$ sector.

\subsection{Observables}
In order to deal with the BRST invariant asymptotic states 
of the $OSp$ invariant string field theory, 
we define the BRST invariant observables corresponding to them. 
They are of the form 
\begin{equation}
\mathcal{O}=\langle ~| \Phi  \rangle~.
\end{equation}
Here $|~ \rangle$ is a first quantized string state and the inner product 
should be considered as including the integrations in the zero-mode part. 
The BRST transformation of this quantity is given as 
\begin{equation}
\delta_{\mathrm{B}}\mathcal{O}
=
\langle~| \Big(Q_{\mathrm{B}}| \Phi\rangle 
                 +g| \Phi *\Phi  \rangle\Big)~.
\end{equation}
In the discussion of the asymptotic sates, 
the second term in the transformation can be ignored.
Therefore for BRST invariant states, we should impose the condition
\begin{equation}
\langle~|Q_{\mathrm{B}} |\Phi \rangle 
=
0~,
\end{equation}
which implies 
\begin{equation}
Q_{\mathrm{B}} |~\rangle =0~.
\end{equation}
For a BRST exact state $|~ \rangle =Q_{\mathrm{B}}|~ \rangle^\prime $, 
\begin{equation}
\mathcal{O}
\simeq
\delta_{\mathrm{B}} {}^\backprime \langle~| \Phi  \rangle~,
\end{equation}
up to multi-string contribution. 
Therefore $|~\rangle$ should be chosen from a nontrivial cohomology class 
of $Q_{\mathrm{B}}$, which was given in 
the previous subsection. 

The string field $|\Phi \rangle$ can be expanded in terms of 
$\pi_0$ and $\bar{\pi}_0$ as
\begin{equation}
|\Phi \rangle 
=
|\bar{\phi}\rangle 
+i\pi_0|\bar{\chi}\rangle +i\bar{\pi}_0|\chi \rangle 
+i\pi_0\bar{\pi}_0|\phi \rangle 
~.
\end{equation}
Substituting this into the kinetic term of the action
eq.(\ref{eq:actionOSp}), 
the only term which is quadratic in $\bar{\phi}$ is  
\begin{equation}
\int dt \int d\alpha \, \frac{d^{26}p}{(2\pi)^{26}}
  \; \frac{1}{2} \langle \bar{\phi} | \bar{\phi} \rangle~.
\end{equation}
The interaction terms are at most quadratic in $\bar{\phi}$. 
Therefore $\bar{\phi}$ can be regarded as an auxiliary field 
and integrated out. 
We identify $\phi$ with the usual physical closed string modes. 
If one integrates $\bar{\phi}$ out,
the kinetic term for $\phi$ looks quite different from
that of the usual field theory. 
It rather looks similar to that of the stochastic quantization. 

Thus for constructing the observables, we discard the case when 
$\langle ~|\Phi\rangle$ is an auxiliary mode. 
Then $|~\rangle$ should be of the form eq.(\ref{brscohomology1}). 
Making the zero-mode integral explicit, 
one can describe the observables constructed above as
\begin{equation}
\mathcal{O}(t,k)
=
\frac{i}{2}
\int_{-\infty}^\infty d\alpha \int d\bar{\pi}_0d\pi_0
 {}_{C,\bar{C}}\langle 0| \otimes
 {}_X\langle \overline{\mathrm{primary}}|
 \Phi (t, \alpha ,\pi_0, \bar{\pi}_0, k)\rangle 
\label{eq:observable}
~. 
\end{equation}
Here the integration measure in eq.(\ref{eq:zeromodemeasure})
is used for the zero-mode integration. 
We normalize the state $|\overline{\mathrm{primary}}\rangle_X$ so that 
\begin{equation}
{}_{X}\langle \overline{\mathrm{primary}} |
         \overline{\mathrm{primary}} \rangle_{X}
 =1~.
\end{equation}
If the state $|\overline{\mathrm{primary}}\rangle_{X}
               \otimes|0\rangle_{C,\bar{C}}\,
               (2\pi)^{26} \delta^{26}(p-k)$
satisfies the relation
\begin{eqnarray}
\lefteqn{
  \left(L_{0}+\tilde{L}_{0}-2 \right)
    |\overline{\mathrm{primary}}\rangle_{X}
    \otimes |0\rangle_{C,\bar{C}}
    \, (2\pi)^{26}\delta^{26} (p-k)}
\nonumber\\
&&
=\left(k^{2}+2i\pi_{0}\bar{\pi}_{0}+M^{2}\right)
  |\overline{\mathrm{primary}}\rangle_{X}
  \otimes |0\rangle_{C,\bar{C}}
   \, (2\pi)^{26} \delta^{26} (p-k)~,
\end{eqnarray}
this state is to be considered as a particle state with mass $M$. 
$\mathcal{O}(t,k)$ is BRST exact unless $k^2+M^2=0$.

\subsection{Free propagator}
We would like to study BRST invariant asymptotic states
using the observables constructed above.
Once the auxiliary field $\bar{\phi}$ is integrated out, 
the action no longer possesses the kinetic term similar to 
that of the usual field theory action. 
Therefore it may seem unlikely that these observables 
correspond to usual particle states. 
However, as we will show, the free propagators corresponding to these 
operators yield propagators for particles propagating in $26$ dimensions. 

Let us consider the observables 
$\mathcal{O}_{r}(t_{r},p_r)$ $(r=1,2)$
which are of the form 
eq.(\ref{eq:observable}) corresponding to
a common primary state, i.e.\
$|\overline{\mathrm{primary}}\rangle_{X}
  \equiv |\overline{\mathrm{primary}}_{1}\rangle_{X}
  =|\overline{\mathrm{primary}}_{2}\rangle_{X}$
and $M \equiv M_{1}=M_{2}$.
We define the two-point function 
\begin{equation}
\left\langle \!\! \left\langle
 \tilde{\mathcal{O}}_1(E_1,p_1)\tilde{\mathcal{O}}_2(E_2,p_2)
\right\rangle \!\! \right\rangle
\equiv
\int dt_1dt_2
 \, e^{iE_1t_1+iE_2t_2}
 \,  \langle \! \langle 0|
         \mathrm{T} \, \mathcal{O}_1(t_1,p_1)\mathcal{O}_2(t_2,p_2)
     |0\rangle \! \rangle ~,
\label{eq:2-pointfunction}
\end{equation}
where $|0\rangle\!\rangle$ denotes the vacuum in the second
quantization.
The lowest order contribution can 
be written by using the Feynman propagator as 
\begin{equation}
\prod_{r=1}^{2}
  \left(
    \frac{i}{2}\int d\alpha_{r} d\bar{\pi}_0^{(r)} d\pi_0^{(r)}
  \right)
\frac{i\delta (1,2)2\pi \delta (E_1+E_2)}
{\alpha_1E_1-p_1^2-M^2-2i\pi_0^{(1)}\bar{\pi}_0^{(1)}+i\epsilon}~,
\label{twogreensfunc}
\end{equation}
where $\delta(1,2)$ is given in eq.(\ref{eq:reflector1.5}).
In the string perturbation theory, 
the propagator corresponds to a cylindrical worldsheet 
and it is calculated as 
\begin{eqnarray}
\lefteqn{
 \frac{i \delta (1,2)2\pi \delta (E_1+E_2)}
{\alpha_1E_1-p_1^2-M^2-2i\pi_0^{(1)}\bar{\pi}_0^{(1)}}
} \nonumber\\
&& =
\int dt_1dt_2
\,e^{iE_1t_1+iE_2t_2}\frac{1}{|\alpha_1|}\delta (1,2)
\left[ \theta (\alpha_1)\theta (t_1-t_2)
e^{-i\frac{t_1-t_2}{\alpha_{1}}
     (p_1^2+2i\pi_0^{(1)}\bar{\pi}_0^{(1)}+M^2)}
\right.  \nonumber
\\
& &
\hspace{15em}
\left.
{} +\theta (\alpha_2)\theta (t_2-t_1)
e^{-i \frac{t_2-t_1}{\alpha_{2}}
    (p_2^2+2i\pi_0^{(2)}\bar{\pi}_0^{(2)}+M^2)}
\right]~.
\label{stringpropagator}
\end{eqnarray}

Let us use this expression to evaluate eq.(\ref{twogreensfunc}).
Substituting eq.(\ref{stringpropagator}) into eq.(\ref{twogreensfunc}),
we obtain 
\begin{eqnarray}
\lefteqn{
\left\langle \!\! \left\langle
\tilde{\mathcal{O}}_1(E_1,p_1)\tilde{\mathcal{O}}_2(E_2,p_2)
\right\rangle \!\! \right\rangle_{\mathrm{free}}
}
\nonumber\\
&&  =
\int dt_1dt_2
   \, e^{iE_1t_1+iE_2t_2}(2\pi )^{26}\delta^{26}(p_1+p_2)
\nonumber
\\
& & \qquad 
\times 
\left[ \theta (t_1-t_2) \frac{i}{2}
\int_0^\infty \frac{d\alpha_1}{\alpha_1}
\int \! d\bar{\pi}_0^{(1)}d\pi_0^{(1)}
 e^{-i\frac{t_1-t_2}{\alpha_1}
      (p_1^2+M^2+2i\pi_0^{(1)}\bar{\pi}_0^{(1)})}
\right.  \nonumber
\\
& &
\qquad \quad \;
 \left. {} + \theta (t_2-t_1)
\frac{i}{2}\int_0^\infty \frac{d\alpha_2}{\alpha_2}
\int \!  d\bar{\pi}_0^{(2)}d\pi_0^{(2)}
 e^{-i\frac{t_2-t_1}{\alpha_2}
        (p_2^2+M^2+2i\pi_0^{(2)}\bar{\pi}_0^{(2)})}
\right].~~~~
\end{eqnarray}
The integrations over $\alpha$, $\pi_{0}$ and $\bar{\pi}_{0}$
can be done as
\begin{eqnarray}
&& \frac{i}{2}
\int_0^\infty \frac{d\alpha_1}{\alpha_1}
\int d\bar{\pi}_0^{(1)}d\pi_0^{(1)}
e^{-i\frac{t_1-t_2}{\alpha_1}
   (p_1^2+M^2+2i\pi_0^{(1)}\bar{\pi}_0^{(1)}-i\epsilon )}
\nonumber\\
&& \quad =
\int_0^\infty \frac{d\alpha_1}{\alpha_1}
  \,   i  \frac{t_1-t_2}{\alpha_1}
e^{-i\frac{t_1-t_2}{\alpha_1}(p_1^2+M^2-i\epsilon )}
=
i\int_0^\infty dt e^{-it(p_1^2+M^2-i\epsilon )}
=
\frac{1}{p_1^2+M^2}~,
\end{eqnarray}
where we introduced $t\equiv \frac{t_1-t_2}{\alpha_1}$. 
Therefore we eventually get
\begin{eqnarray}
\lefteqn{
\left\langle \!\!
\left\langle
\tilde{\mathcal{O}}_1(E_1,p_1)\tilde{\mathcal{O}}_2(E_2,p_2)
\right\rangle \!\! \right\rangle_{\mathrm{free}}
}
\nonumber\\
&&=
\int dt_1dt_2e^{iE_1t_1+iE_2t_2}(2\pi )^{26}\delta^{26}(p_1+p_2)
\left[
  \frac{\theta (t_1-t_2)}{p_1^2+M^2}+\frac{\theta (t_2-t_1)}{p_2^2+M^2}
\right]
\nonumber
\\
&& =
\frac{(2\pi )^{26}\delta^{26}(p_1+p_2)}{p_1^2+M^2}
2\pi \delta (E_1)2\pi \delta (E_2)~.
\end{eqnarray}
The reason why we have factors $2\pi \delta (E_r)$
can be understood as follows. 
As we can see from the expression eq.(\ref{stringpropagator}), 
$\frac{L_0+\tilde{L}_0-2}{\alpha}$ is the Hamiltonian on the worldsheet. 
Since this is a BRST exact operator, 
$\mathcal{O}(t+\delta t,p)$ and $\mathcal{O}(t,p)$ are BRST equivalent 
and only the constant mode with $E=0$ survives. 
Thus we here choose 
\begin{eqnarray}
\varphi (p)
&\equiv& 
\int \frac{dE}{2\pi}\tilde{\mathcal{O}}(E,p)
\nonumber
\\
&=&
\mathcal{O}(t=0,p)~,
\label{mathcalO}
\end{eqnarray}
as a representative of these equivalent operators. 

Hence we have 
\begin{equation}
\left\langle \!
\left\langle
\varphi_1(p_1)\varphi_2(p_2)
\right\rangle \! \right\rangle_{\mathrm{free}}
=
\frac{(2\pi )^{26}\delta^{26}(p_1+p_2)}{p_1^2+M^2}~.
\label{freepropagator}
\end{equation}
This coincides with the Euclidean propagator for a particle with mass $M$. 
Thus we have shown that although the string field action possesses an unusual
form, 
modes corresponding to the operators
\begin{eqnarray}
\varphi (p)
&=&
\frac{i}{2}
\int_{-\infty}^\infty d\alpha \int d\bar{\pi}_0 d\pi_0
 \; {}_{C,\bar{C}}\langle 0| \otimes
 {}_X\langle \overline{\mathrm{primary}}|
 \Phi (t=0, \alpha ,\pi_0, \bar{\pi}_0, p)\rangle 
\nonumber
\\
&=&
\frac{i}{2}\int\frac{dE}{2\pi}
\int_{-\infty}^\infty d\alpha \int d\bar{\pi}_0d\pi_0
 \; {}_{C,\bar{C}}\langle 0| \otimes
 {}_X\langle \overline{\mathrm{primary}}|
 \tilde{\Phi} (E, \alpha ,\pi_0, \bar{\pi}_0, p)\rangle ~,
\end{eqnarray}
yield usual propagators. 
Here $\tilde{\Phi}$ is the Fourier transform of $\Phi$ with respect to $t$. 
$\varphi(p)$ corresponds to a particle included in string theory. 

For other modes, things are not so simple in general. 
For our purpose, it is necessary to check 
the two-point functions involving BRST exact observables 
$\mathcal{O}=\langle ~|Q_{\mathrm{B}}|\Phi \rangle$. 
Notice that the free propagator
\begin{equation}
\Big\langle \!\!
\Big\langle  \,
\langle ~|Q_{\mathrm{B}}|\Phi \rangle ~{}^\backprime \langle~|\Phi \rangle 
\,
\Big\rangle \!\! \Big\rangle_{\mathrm{free}}~,
\label{null}
\end{equation}
is not necessarily $0$, even if $Q_{\mathrm{B}}|~\rangle^\prime =0$. 
Indeed, calculating this quantity boils down to evaluating
\begin{equation}
\Big(
\langle ~|Q_{\mathrm{B}}
\Big)
\exp \left( -it\frac{L_0+\tilde{L}_0-2-i\epsilon}{\alpha} \right)
|~\rangle^\prime
\end{equation}
which is $0$ if we can make $Q_{\mathrm{B}}$ act on the state on the right. 
In order to do so, we should perform a partial integration over $\alpha$. 
If the integrand does not vanish for $\alpha \rightarrow \infty$,
we have a nonvanishing surface term and obtain a nonvanishing
result.\footnote{Notice that for the BRST exact operator 
$\mathcal{O}(t+\delta t,p)-\mathcal{O}(t,p)$ 
we discussed above, this does not happen.} 
For example, $\varphi (p)$ itself is actually BRST exact for 
$p^2+M^2\neq 0$, but we have eq.(\ref{freepropagator}). 
Anyway, contributions for the correlation functions involving such 
BRST exact operators come from the boundary of the moduli space of the 
worldsheet, which is usual in string perturbation theory. 
Therefore, we do not expect to find particle poles such as
$\frac{1}{p^2+M^2}$ in such correlators. 
Indeed it is straightforward to check that the correlation function 
eq.(\ref{null}) with $Q_\mathrm{B}|~\rangle^\prime =0$ 
does not yield such poles, provided $|~\rangle$ exists for $p^2+M^2=0$. 

Now that we identify the modes of $\Phi$ corresponding to the particle states 
in string theory, we can construct the asymptotic states using them. 
Wick rotating as $x^{26}\rightarrow x^0=-ix^{26}$, we can canonically quantize 
the theory considering $x^0$ as time. 
Since the free propagator corresponding to $\varphi (p)$ coincides with 
that for a particle with mass $M$, it is straightforward to define 
properly normalized asymptotic states using these operators. 

We may be able to proceed and calculate the S-matrix elements for 
these asymptotic states. 
The calculations will be essentially the same as those in 
\cite{Kugo:1987rq}. 
Using a generalization of the Parisi-Sourlas formula, 
we may be able to show that the S-matrix elements coincide with 
those of the light-cone gauge string field theory. 
However, in this paper we will rather calculate the correlation functions 
of the observables and define the S-matrix elements using them. 
By doing so, we can proceed without Wick rotating the $\pm$ directions. 
Such a Wick rotation is necessary for deriving the Parisi-Sourlas type 
formula in this context\cite{Neveu:1986cu}\cite{Kugo:1987rq}.

\section{Correlation Functions and S-matrix Elements}
\label{sec:S-matrix-equivalence}

We have one observable $\varphi (p)$ for one primary state in the 
Hilbert space of $X^\mu$. 
These primary states are in one-to-one correspondence to the states with 
physical polarizations in string theory. 
The correlation functions of the operators
$\varphi (p)$ can be considered as those in 
the $26$ dimensional Euclidean space. 
Essentially, what we would like to show in this section is 
that these correlation functions can be considered as 
the correlation functions for 
the bosonic string field theory. 
We will prove that the S-matrix elements 
derived from these correlation functions 
coincide with those of the light-cone gauge string field theory.

\subsection{Correlation functions}


Let us consider $N$-point correlation functions ($N\geq 3$)
\begin{equation}
\bigg\langle \!\! \bigg\langle
 \prod_{r=1}^N\varphi_r (p_r)
\bigg\rangle \!\! \bigg\rangle ~,
\end{equation}
of the observables 
\begin{equation}
\varphi_{r}(p_r)
=
\int \frac{dE_r}{2\pi}\tilde{\mathcal{O}}_r (E_r,p_r)
\qquad (r=1,\ldots, N)~,
\end{equation}
which are 
made from Virasoro primary states 
$|\overline{\mathrm{primary}}_r\rangle_X$ 
corresponding to particles with mass $M_r$.
We will show that these correlation functions yield S-matrix elements
for string theory. 
In order to evaluate this, we start from the correlation function 
\begin{equation}
\bigg\langle \!\! \bigg\langle
 \prod_{r=1}^N\tilde{\mathcal{O}}_r (E_r,p_r)
\bigg\rangle \!\! \bigg\rangle
\equiv  \prod_{r=1}^N 
\left(
\int dt_re^{iE_rt_r}
\right)
\langle \! \langle 0| \, \mathrm{T}
   \prod_{r=1}^N\mathcal{O}_r (t_r,p_r) \, |0\rangle \! \rangle~.
 \label{eq:NptGreen}
\end{equation}

The $OSp$ invariant string field theory can be regarded
as a light-cone gauge string field theory with
``transverse" space-time coordinates
$X^M$ and the perturbative expansion can be obtained
as in the usual light-cone gauge string field theory. 
The correlation function (\ref{eq:NptGreen})
can be calculated perturbatively and takes the form
\begin{eqnarray}
\bigg\langle \!\! \bigg\langle
 \prod_{r=1}^N\tilde{\mathcal{O}}_r (E_r,p_r)
\bigg\rangle \!\! \bigg\rangle
&=&
\prod_{r=1}^N \left(
  \frac{i}{2}\int d\alpha_r d\bar{\pi}_0^{(r)}d\pi_0^{(r)}
\frac{i}
{\alpha_r E_r-p_r^2-M_r^2-2i\pi_0^{(r)}\bar{\pi}_0^{(r)}} \right)
\nonumber
\\
& &
{}\times 
\delta^{OSp}\bigg(\sum_{s=1}^N p_s^{OSp}\bigg)
 \, G_{\mathrm{amputated}}^{OSp}(p_1^{OSp},\cdots ,p_N^{OSp})~.
\label{ngreensfunc}
\end{eqnarray}
Here $p_r^{OSp}$ denotes the zero-modes
$p_r^{OSp}=(E_r,\alpha_r,p_r^\mu ,\bar{\pi}_0^{(r)},\pi_0^{(r)})$
of the $r$-th string and
\begin{equation}
\delta^{OSp}\bigg( \sum_{r=1}^Np_r^{OSp}\bigg)
=
2\pi \delta \bigg(\sum_{r=1}^NE_r \bigg)
\delta \bigg(\sum_{r=1}^N\alpha_r\bigg)
(2\pi )^{26}\delta^{26}\bigg( \sum_{r=1}^Np_r \bigg)
2i \bigg( \sum_{r=1}^N \bar{\pi}_0^{(r)} \bigg)
    \bigg( \sum_{r=1}^N \pi_0^{(r)} \bigg)~.
\end{equation}
$G_{\mathrm{amputated}}^{OSp}(p_1^{OSp},\cdots ,p_N^{OSp})$ 
is the amputated Green's function,
which is expressed as
\begin{eqnarray}
\lefteqn{
G_{\mathrm{amputated}}^{OSp}(p_1^{OSp},\cdots ,p_N^{OSp})
}
\nonumber\\
&&
=
\sum_G
\int \prod_Id\alpha_I\prod_nd(t_n-t_{n-1})
F_G(p_1^{OSp},\cdots ,p_N^{OSp};\alpha_I,t_n-t_{n-1})~,
\label{Gtrunc}
\end{eqnarray}
where $G$ denotes a light-cone string diagram, 
$\alpha_I$ denotes the string length for an internal line of the 
diagram and $t_n$ denotes the proper time for a three-string vertex. 
Because of the conservation of the string length,
some of $\alpha_I$ can be expressed by 
other $\alpha$'s through delta functions involved in $F_G$. 
Independent $\alpha_I$'s and $t_n-t_{n-1}$ can be regarded as the 
the moduli of the Riemann surface corresponding to
the light-cone string diagram. 
If none of the ratios 
$\frac{\alpha_I}{\alpha_{I^\prime}},\frac{t_n-t_{n-1}}{\alpha_I}$ 
are $0$ or infinity, 
the diagram corresponds to a nondegenerate Riemann surface. 
Using the expression eq.(\ref{Gtrunc}), 
one can rewrite eq.(\ref{ngreensfunc}) as 
\begin{equation}
\bigg\langle \!\! \bigg\langle
 \prod_{r=1}^N \tilde{\mathcal{O}}_r(E_r,p_r)
 \bigg\rangle \!\! \bigg\rangle
=
\sum_G
\int \prod_Id\alpha_I\prod_nd(t_n-t_{n-1})
I_G(p_1^{OSp},\cdots ,p_N^{OSp};\alpha_I,t_n-t_{n-1})~,
\end{equation}
where
\begin{eqnarray}
\lefteqn{
I_G(p_1^{OSp},\cdots ,p_N^{OSp};\alpha_I,t_n-t_{n-1})
} \nonumber\\
&& =
\prod_{r=1}^N \left(
  \frac{i}{2}\int d\alpha_r d\bar{\pi}_0^{(r)}d\pi_0^{(r)}
\frac{i}
{\alpha_r E_r-p_r^2-M_r^2-2i\pi_0^{(r)}\bar{\pi}_0^{(r)}}
\right)
\nonumber
\\
& & \quad
{} \times 
\delta^{OSp} \bigg(\sum_{s=1}^N p_s^{OSp} \bigg)
F_G(p_1^{OSp},\cdots ,p_N^{OSp};\alpha_I,t_n-t_{n-1})~.
\label{integrand}
\end{eqnarray}

In $I_G$, the moduli concerning the external lines are already integrated. 
In usual treatment of string theory, 
the moduli concerning the external lines are taken care of
first\footnote{Usually we replace them by local vertex operators. }
and string amplitudes are given as integrations over the moduli space 
of the rest of the worldsheet. 
In doing so, one tacitly discards contributions from degenerate surfaces 
which may have some physical significance.
For example, masses of massive string states are expected to be shifted
by radiative corrections. 
However such effects are not included in defining the 
S-matrix of string theory,
because the moduli of the external line propagators should 
be treated simultaneously with the other moduli
in order to study these effects. 
Since we would like to relate our correlation functions
to the results of usual formulation of bosonic string theory,
we will follow the same order. 
We will calculate the integrand $I_G$ for diagrams corresponding 
to nondegenerate Riemann surfaces. 

Let us first show that for $p_r^2+M_r^2\sim 0$, $I_G$ behaves as
\begin{equation}
I_G=\frac{C}{p_r^2+M_r^2}+\mbox{less singular terms}~. 
\end{equation}
Suppose $r\neq N$ for example. 
We can integrate over
$\alpha_N, \pi_0^{(N)}, \bar{\pi}_0^{(N)}$ in eq.(\ref{integrand})
and we obtain
\begin{eqnarray}
\lefteqn{
I_G(p_1^{OSp},\cdots ,p_N^{OSp};\alpha_I,t_n-t_{n-1})
} \nonumber\\
&& =
\prod_{r=1}^{N-1}
 \left( \frac{i}{2}\int d\alpha_r d\bar{\pi}_0^{(r)}d\pi_0^{(r)}
   \frac{i}{\alpha_rE_r-p_r^2-M_r^2-2i\pi_0^{(r)}\bar{\pi}_0^{(r)}}
 \right)
\nonumber\\
&&
\quad {} \times
2\pi \delta \bigg(\sum_{s=1}^N E_s\bigg)
(2\pi )^{26}\delta^{26}\bigg(\sum_{s=1}^N p_s\bigg)
\;
F_G \bigg(p_1^{OSp},\cdots ,p_{N-1}^{OSp},-\sum_{s=1}^{N-1}p_s^{OSp};
    \alpha_I,t_n-t_{n-1}\bigg)
\nonumber \\
&&
\quad {}\times 
\frac{i}
{-E_N\sum_{s=1}^{N-1}\alpha_s-p_N^2-M_N^2
-2i\bigg(\sum_{s=1}^{N-1}\pi_0^{(s)} \bigg)
   \bigg(\sum_{s=1}^{N-1}\bar{\pi}_0^{(s)}\bigg)}~.
\end{eqnarray}
We are interested in the singular behavior of this quantity
at $p_r^2+M_r^2=0$. 
Since $F_G$ is given by a product of factors 
$e^{-i \frac{t}{\alpha}(p^2+2i\pi_0\bar{\pi}_0+M^2)}$ and those from 
the three-string vertices, it cannot be singular at $p_r^2+M_r^2=0$. 
Therefore, for generic momenta $p_r^\mu$, such singularities come from the 
integration over $\alpha_r$. 
For integrating over
$\alpha_r, \pi_0^{(r)}, \bar{\pi}_0^{(r)}~(r=1,\cdots ,N-1)$, 
we rewrite the propagator again as 
\begin{eqnarray}
&&
\frac{i}
{\alpha_r E_r-p_r^2-M_r^2-2i\pi_0^{(r)}\bar{\pi}_0^{(r)}}
 \\
&&=
\int dt\, e^{iE_r t}\frac{1}{|\alpha_r|}
\left[
 \theta (\alpha_r)\theta (t)
 \,e^{-i\frac{t}{\alpha_{r}}
       (p_r^2+2i\pi_0^{(r)}\bar{\pi}_0^{(r)}+M_r^2)}
 +\theta (-\alpha_r)\theta (-t)
\, e^{-i\frac{t}{\alpha_{r}}(p_r^2+2i\pi_0^{(r)}\bar{\pi}_0^{(r)}+M_r^2)}
\right]. \nonumber
\end{eqnarray}
Thus the integrations we should perform are of the form 
\begin{equation}
\frac{i}{2}
\int_0^\infty \frac{d\alpha_r}{\alpha_r}
\int d\bar{\pi}_0^{(r)}d\pi_0^{(r)}
\, e^{-i\frac{t}{\alpha_{r}}
     (p_r^2 + 2i\pi_0^{(r)}\bar{\pi}_0^{(r)}+M_r^2)}
\, f(\alpha_r,\pi_0^{(r)},\bar{\pi}_0^{(r)},p_r)~. 
\label{final}
\end{equation}

The singular behavior of the integral eq.(\ref{final}) 
is related to the behavior of the integrand around $\alpha_r\sim 0$. 
If we take $\alpha_r\rightarrow 0$ keeping the light-cone string 
diagram $G$ nondegenerate, 
the $r$-th external line can be replaced by a local vertex operator
times some 
factor depending on $\alpha_r$. 
{}From the form of the three-string vertex, one can deduce
\begin{eqnarray}
F_G(p_r^{OSp};\alpha_I,t_n-t_{n-1})
&=&
\left.
\alpha_r^{p_r^2+2i\pi_0^{(r)}\bar{\pi}_0^{(r)}+M_r^2}
\, F_G(p_r^{OSp};\alpha_I,t_n-t_{n-1})
\right|_{p_r^2+M_r^2=\alpha_r=\pi_0^{(r)}=\bar{\pi}_0^{(r)}=0}
\nonumber
\\
& &
+
\mbox{higher order terms}
\end{eqnarray}
for $\alpha_r \sim 0$.\footnote{Here,
$\left.{}\right|_{p_r^2+M_r^2=\alpha_r=\pi_0^{(r)}=\bar{\pi}_0^{(r)}=0}$ 
should be understood so that we put $p_r^2+M_r^2$ first
and then take the other variables to be $0$.}

Therefore, eq.(\ref{final}) can be evaluated as 
\begin{equation}
\left.
\frac{1}{p_r^2+M_r^2}
\, f(\alpha_r,\pi_0^{(r)},\bar{\pi}_0^{(r)},p_r)
\right|_{p_r^2+M_r^2=\alpha_r=\pi_0^{(r)}=\bar{\pi}_0^{(r)}=0}
+
\mbox{less singular terms}~.
\end{equation}
Here ``less singular terms" indicates
terms with less singular behavior at 
$p_r^2+M_r^2=0$. 
Thus we can see that $I_G$ has at most a simple pole at $p_r^2+M_r^2=0$.

We consider this pole as the one for a particle with mass $M_r$ and 
deduce S-matrix elements. 
In order to do so, we should investigate the most singular part of 
$I_G$ for $p_r^2+M_r^2 \sim  0~(r=1,2,\cdots ,N)$. 
This can be done by successively integrating 
over $\alpha_r,\pi_0^{(r)},\bar{\pi}_0^{(r)}$ 
for $r=1,\cdots ,N-1$. 
Calculations are essentially the same as above and we eventually get
\begin{eqnarray}
\lefteqn{
I_G(p_1^{OSp},\cdots ,p_N^{OSp};\alpha_I,t_n-t_{n-1})
} \nonumber\\
&& =
-i \left(\prod_{r=1}^N\frac{2\pi \delta (E_r)}{p_r^2+M_r^2}\right)
(2\pi )^{26}\delta^{26}\bigg(\sum_{r=1}^N p_r \bigg)
\, \left. F_G(p_r^{OSp};\alpha_I,t_n-t_{n-1})
\rule{0em}{2.5ex}
\right|_{p_r^2+M_r^2=\alpha_r=\pi_0^{(r)}=\bar{\pi}_0^{(r)}=0}
\nonumber
\\
& &
\quad {}+
\mbox{less singular terms}~.
\label{IG}
\end{eqnarray}
On the course of this calculation,
we encounter higher order poles of $p_N^2+M_N^2$. 
These should cancel with each other, because $I_G$ can have
at most simple poles at $p_r^2+M_r^2=0$ as we have shown above.

\subsection{S-matrix elements}
Substituting eq.(\ref{IG}) into eq.(\ref{integrand}), 
we formally obtain 
\begin{eqnarray}
\lefteqn{
\bigg\langle \!\! \bigg\langle
 \prod_{r=1}^N\tilde{\mathcal{O}}_r(E_r,p_r)
\bigg\rangle \!\! \bigg\rangle
}\nonumber\\
&& =  -i
\left(\prod_{r=1}^N\frac{2\pi \delta (E_r)}{p_r^2+M_r^2}\right)
(2\pi )^{26}\delta^{26} \bigg( \sum_{r=1}^N p_r \bigg)
 \left. G_{\mathrm{amputated}}^{OSp}(p_r^{OSp})
\rule{0em}{2.5ex}
\right|_{p_r^2+M_r^2=\alpha_r=\pi_0^{(r)}=\bar{\pi}_0^{(r)}=0}
\nonumber
\\
& & \quad
{}+\mbox{less singular terms}~.
\end{eqnarray}
Hence the correlation function for $\varphi_{r} (p_{r})$
becomes
\begin{eqnarray}
\lefteqn{
\bigg\langle \!\! \bigg\langle
 \prod_{r=1}^N \varphi_{r}(p_r)
\bigg\rangle \!\! \bigg\rangle}
\nonumber\\
&& =
\left(\prod_{r=1}^N \frac{1}{p_r^2+M_r^2} \right)
(-i)(2\pi )^{26}\delta^{26} \bigg(\sum_{r=1}^N p_r \bigg)
\left. G_{\mathrm{amputated}}^{OSp}(p_r^{OSp})
\rule{0em}{2.5ex}
\right|_{p_r^2+M_r^2=E_r=\alpha_r=\pi_0^{(r)}=\bar{\pi}_0^{(r)}=0}
\nonumber
\\
& & \quad 
{} +\mbox{less singular terms}~.
\label{reductionformula}
\end{eqnarray}
Considering this correlation function as a correlation function of a 
Euclidean $26$ dimensional field theory, 
we can Wick rotate it and derive the S-matrix element.  
Let us define the Lorentzian momentum
\begin{equation}
p^\mathrm{L}=(p_0,p_1,\cdots ,p_{25})~,
\end{equation}
with $p_0=ip_{26}$. 
The S-matrix element $S(p_r^\mathrm{L})$ we obtain is
\begin{equation}
S(p_r^\mathrm{L})
=
(2\pi )^{26}\delta \Big(\sum_{r=1}^{N} p_r^\mathrm{L} \Big)
\left. G_{\mathrm{amputated}}^{OSp}(p_r^{OSp})
\rule{0em}{2.5ex}
\right|_{p_r^2+M_r^2=E_r=\alpha_r=\pi_0^{(r)}=\bar{\pi}_0^{(r)}=0}~.
\label{Smatrixelement}
\end{equation}

We would like to show that this S-matrix element coincides with that 
in the light-cone gauge string field theory. 
In order to do so, it is convenient to express $S(p_r^\mathrm{L})$ 
by using the S-matrix elements for
the $OSp$ invariant string field theory.

\subsubsection*{S-matrix elements for the $OSp$ invariant string field theory}
If we do not care about the BRST symmetry,
the $OSp$ invariant string field theory is 
just a light-cone gauge string field theory on a flat supermanifold and 
the S-matrix elements can be defined in the usual way. 
Since the $OSp$ invariant string field theory possesses
the $OSp(27,1|2)$ symmetry,
the S-matrix elements are invariant under this symmetry. 
An on-shell particle state is specified by its momentum $p^{OSp}$
and polarization $\epsilon^{OSp}$. 
As functions of these variables,
the S-matrix elements $S^{OSp}$ can be expressed as
\begin{eqnarray}
S^{OSp} \left( p^{OSp}_r,\epsilon^{OSp}_r \right)= 
     \delta^{OSp} \Big( \sum_r p^{OSp}_r \Big)
f^{OSp} \left( p^{OSp}\cdot p^{OSp}\, ,\,
         p^{OSp}\cdot \epsilon^{OSp}\, ,\,
         \epsilon^{OSp} \cdot \epsilon^{OSp} \right)~, 
\label{eq:S-OSp}
\end{eqnarray}
where $p^{OSp}\cdot p^{OSp}\, ,\,
         p^{OSp} \cdot \epsilon^{OSp}$ 
and $\epsilon^{OSp} \cdot \epsilon^{OSp}$
denote the general invariants of the $OSp(27,1|2)$ group
composed of momenta and polarizations.

By definition, 
$f^{OSp}$ coincides with $G_{\mathrm{amputated}}^{OSp}$
with all the momenta on-shell:
\begin{equation}
f^{OSp}=
\left. G_{\mathrm{amputated}}^{OSp}
\rule{0em}{2.5ex}
\right|_{p_r^2+M_r^2=0}~.
\label{Gtrunc=f}
\end{equation}
The oscillator parts 
$|0\rangle_{C,\bar{C}}\otimes |\overline{\mathrm{primary}}_r\rangle_X $ 
of the states under consideration correspond to polarizations 
whose components involving $\pm ,C,\bar{C}$ indices are zero. 
Therefore the last factor on the right hand side of eq.(\ref{Smatrixelement}) 
can be given by $f^{OSp}$ with $p^{OSp},\epsilon^{OSp}$
whose components involving $\pm ,C,\bar{C}$ indices are zero. 

The S-matrix element $S(p_r^\mathrm{L})$ in eq.(\ref{Smatrixelement}) 
can be considered as a function of these polarizations. 
Such polarizations can be considered as tensors in $26$ dimensions. 
Let us denote the Lorentzian version of these tensors by $\epsilon^\mathrm{L}$, 
and the S-matrix element as a function of these polarizations by 
$S(p_r^\mathrm{L},\epsilon_r^\mathrm{L})$. 
The $OSp$ invariant combinations 
$p^{OSp}\cdot p^{OSp}$, $p^{OSp}\cdot \epsilon^{OSp}$,
$\epsilon^{OSp}\cdot \epsilon^{OSp}$ 
for $p^{OSp}$, $\epsilon^{OSp}$ satisfying the above-mentioned conditions 
coincide with the $SO(25,1)$ invariant combinations
$p^\mathrm{L}\cdot p^\mathrm{L}$,
$p^\mathrm{L}\cdot \epsilon^\mathrm{L}$,
$\epsilon^\mathrm{L}\cdot \epsilon^\mathrm{L}$ 
which are defined in an obvious way following the $OSp$ version. 
Hence the S-matrix element $S(p_r^\mathrm{L},\epsilon_r^\mathrm{L})$ 
can be written by using
eqs.(\ref{Smatrixelement}) and (\ref{Gtrunc=f}) as
\begin{eqnarray}
S(p_r^\mathrm{L},\epsilon_r^\mathrm{L})
&=&
(2\pi )^{26}\delta^{26}\Big( \sum_r p_r^\mathrm{L} \Big)
f^{OSp}(p^{OSp}\cdot p^{OSp} \, , \, p^{OSp}\cdot \epsilon^{OSp}
        \, , \, \epsilon^{OSp} \cdot \epsilon^{OSp}) 
\nonumber
\\
&=&
(2\pi )^{26}\delta^{26}\Big( \sum_r p_r^\mathrm{L} \Big)
f^{OSp}(p^\mathrm{L}\cdot p^\mathrm{L} \, , 
         \, p^\mathrm{L}\cdot \epsilon^\mathrm{L}
        \, , \, \epsilon^\mathrm{L} \cdot \epsilon^\mathrm{L})~.
\label{S=OSp}
\end{eqnarray}

\subsubsection*{Light-cone gauge string field theory}
By construction $f^{OSp}$ is related to an S-matrix element of the 
light-cone gauge string field theory. 
Using this fact, one can deduce from eq.(\ref{S=OSp}) that 
the S-matrix element 
$S(p_r^\mathrm{L},\epsilon_r^\mathrm{L})$ coincides with that 
of the light-cone gauge string field theory, as was done in \cite{Kugo:1987rq}. 
Here, for completeness, we will give a proof of this fact. 

The light-cone gauge string field theory we have in mind has the 
action 
\begin{eqnarray}
\lefteqn{
\int dt \left[
  \frac{1}{2}
   \int d1_{\mathrm{LC}}d2_{\mathrm{LC}}\, 
   {}^{\mathrm{LC}}\left\langle R(1,2) \left|\Phi\right\rangle_{1}
               \right.
       \left( i\frac{\partial}{\partial t}
              -\frac{L_0^{\mathrm{LC}(2)}+\tilde{L}_0^{\mathrm{LC}(2)}-2}
                    {\alpha_2}
       \right)  \left|\Phi\right\rangle_{2}
   \right.} \hspace{3.5em}\nonumber\\
&& \left.
{} + \frac{2g}{3}
\int d1_{\mathrm{LC}}
     d2_{\mathrm{LC}}
     d3_{\mathrm{LC}}
      \, {}^{\mathrm{LC}}\left\langle V_3^0(1,2,3)\right|
\Phi\rangle_1|\Phi\rangle_2|\Phi\rangle_3
\right]~,
\label{eq:actionLC}
\end{eqnarray}
where $dr_{\mathrm{LC}}$ is defined as
\begin{equation}
dr_{\mathrm{LC}}
\equiv
 \frac{\alpha_rd\alpha_r}{2}
 \frac{d^{24}p_r}{(2\pi)^{25}}~,
 \label{eq:zeromodemeasureLC}
\end{equation}
$L_0^{\mathrm{LC}}$ and $\tilde{L}_0^{\mathrm{LC}}$
 are the Virasoro generators 
for the light-cone variables and
${}^{\mathrm{LC}}\left\langle R(1,2) \right|$ 
and ${}^{\mathrm{LC}}\left\langle V_3^0(1,2,3)\right|$ are defined
in appendix \ref{sec:conventions}.

Now let us consider the S-matrix elements 
of this light-cone gauge string field theory 
for the external states with momenta and polarizations 
$(p_r^\mathrm{L},\epsilon_r^\mathrm{L})~(r=1,\cdots ,N)$. 
The light-cone gauge string field theory possesses
the $O(25,1)$ Lorentz symmetry and 
the S-matrix elements 
$S^{\mathrm{LC}}$ take the form
\begin{equation}
S^{\mathrm{LC}}(p_r^\mathrm{L},\epsilon_r^\mathrm{L} ) =
     (2\pi)^{26}\delta^{26}\Big( \sum_r p_r^\mathrm{L}\Big)
     f^{\mathrm{LC}}
        (p^\mathrm{L}\cdot p^\mathrm{L} \, ,\,
         p^\mathrm{L}\cdot \epsilon^\mathrm{L} \, ,\,
         \epsilon^\mathrm{L} \cdot \epsilon^\mathrm{L} )~,
\label{eq:S-LC}
\end{equation}
where $p^\mathrm{L}\cdot p^\mathrm{L}\,$,
         $p^\mathrm{L} \cdot \epsilon^\mathrm{L} \,$ 
and $\epsilon^\mathrm{L} \cdot \epsilon^\mathrm{L}$
denote the general invariants of the $O(25,1)$ group. 

The $OSp$ invariant string field theory
is constructed from the light-cone gauge string field theory
through the $OSp$ extension. 
On the worldsheet, the $OSp$ extension corresponds to adding two 
bosons $X^{25},X^{26}$ and two fermions $C,\bar{C}$ with spin $0$. 
Therefore it is conceivable that the $OSp$ extended theory yields 
the same results as those from the original theory 
in some situations. 
Indeed, one can prove that for momenta $p^{OSp}$ 
and polarizations $\epsilon^{OSp}$ whose components 
involving $25,26 ,C,\bar{C}$ indices 
are zero, 
\begin{equation}
f^{OSp}
(p^{OSp}\cdot p^{OSp},p^{OSp}\cdot \epsilon^{OSp},
   \epsilon^{OSp}\cdot \epsilon^{OSp})
=
f^\mathrm{LC}
(p^{OSp}\cdot p^{OSp},p^{OSp}\cdot \epsilon^{OSp},
   \epsilon^{OSp}\cdot \epsilon^{OSp}).
\label{OSp=LC}
\end{equation}

The proof goes as follows. 
$f^{OSp}$ is calculated using the perturbation theory 
of strings as in eq.(\ref{Gtrunc}). 
On the worldsheet we have variables $X^\mu ,C,\bar{C}$.
It is easy to see that 
the part of the worldsheet theory consisting of bosons 
$X^{25},X^{26}$ and fermions $C,\bar{C}$
with spin $0$ can be considered as a topological field theory. 
Indeed if we define $Z=X^{25}+iX^{26}$, the $OSp$ generator $M^{ZC}$
is nilpotent and can be considered as a BRST operator.
The Hamiltonian is BRST exact and the three-string vertex is BRST invariant. 
The states whose momenta and polarizations do not have components
involving $25,26 ,C,\bar{C}$ indices are invariant under this BRST symmetry. 
Therefore contributions of these variables to the 
integrand in eq.(\ref{Gtrunc}) do not depend on the moduli.  
Thus we calculate them on the surface where $t_n-t_{n-1}\rightarrow\infty$. 
Then we get the factor
\begin{equation}
\lim_{T\to \infty} \exp\left[ -\frac{T}{|\alpha |}\left(
          p_{25}^2+p_{26}^2
                     +2i\pi_0\bar{\pi}_{0}\right)\right]
=2\pi i \delta\left( p_{25}\right)
        \delta\left( p_{26}\right)
        \bar{\pi}_0\pi_0~,
\label{eq:regularization-delta}
\end{equation}
from the propagator part. 
Hence only the state
$|0\rangle_{C,\bar{C}}\otimes |0\rangle_{X^{25},X^{26}}$ 
with $p_{25}=p_{26}=\pi_0=\bar{\pi}_0=0$ contribute to the amplitudes. 
Using these, we can show that the contributions from these variables
are $1$ and we are left with the light-cone gauge string amplitudes
derived from the action eq.(\ref{eq:actionLC}). 

The conditions satisfied by the $p^{OSp},\epsilon^{OSp}$
in eq.(\ref{OSp=LC}) is related to those satisfied by
$p^{OSp},\epsilon^{OSp}$ in eq.(\ref{S=OSp}) 
via Wick rotations and space rotations. 
Since $f^{OSp},f^\mathrm{LC}$ depend only on the combinations
invariant under such manipulations, 
eq.(\ref{OSp=LC}) can be used to replace $f^{OSp}$ in eq.(\ref{S=OSp})
by $f^\mathrm{LC}$ and we finally obtain
\begin{equation}
S(p_r^\mathrm{L},\epsilon_r^\mathrm{L})
=
S^\mathrm{LC}(p_r^\mathrm{L},\epsilon_r^\mathrm{L}).
\end{equation}

Before closing this section, one comment is in order. 
The left hand side of eq.(\ref{eq:regularization-delta})
can be considered to give a regularization of the delta function
on the right hand side, which preserves various symmetries. 
Using this regularization, we obtain
\begin{equation}
\left.
2\pi i   \delta\left( p_{25}\right)
        \delta\left( p_{26}\right)
        \bar{\pi}_0\pi_0
        \rule{0em}{1em}\right|_{p=\pi_0=\bar{\pi}_0=0}
=
1~.
\end{equation}
Wick rotating this, we can see that eq.(\ref{reductionformula}) 
can be rewritten as
\begin{eqnarray}
\lefteqn{
\bigg\langle \!\! \bigg\langle
 \prod_{r=1}^N \varphi_{r}(p_r)
\bigg\rangle \!\! \bigg\rangle
} \nonumber\\
&&=
\left(\prod_{r=1}^N \frac{1}{p_r^2+M_r^2} \right)
\left.\left[
\delta^{OSp} \bigg(\sum_{r=1}^N p_r^{OSp} \bigg)
 G_{\mathrm{amputated}}^{OSp}(p_r^{OSp})
\right]
\right|_{p_r^2+M_r^2=E_r=\alpha_r=\pi_0^{(r)}=\bar{\pi}_0^{(r)}=0}
\nonumber
\\ 
& & \quad 
{} +\mbox{less singular terms}~,
\end{eqnarray}
in which form the relation to the Parisi-Sourlas reduction may 
be clearer.

\section{Discussion}
\label{sec:discussions}

In this paper,
we have defined BRST invariant observables
in the $OSp$ invariant string field theory
and evaluated correlation functions of them.
We have shown that the S-matrix elements derived from
these correlation functions coincide with those
of the light-cone gauge string field theory.

Our results will be useful to understand the structure 
of the $OSp$ invariant string field theory and 
explore the proposal in \cite{Baba:2006rs} further. 
The BRST invariant solitonic operators constructed in 
\cite{Baba:2006rs} may be regarded as another kind of 
observables besides those we constructed in this paper. 
In this paper, we only care about the particle poles 
and observables are only BRST invariant up to terms 
nonlinear in the string fields. 
Since the boundary states are off-shell states, 
the observables involving these states should be 
BRST invariant taking the nonlinear terms into account. 
What we proposed in \cite{Baba:2006rs} are  
such observables. 

The solitonic operators in \cite{Baba:2006rs} 
correspond to D-branes or ghost D-branes \cite{Okuda:2006fb}. 
However since we only calculated cylinder amplitudes, 
we were not able to distinguish the two. 
Using the method developed in this paper, we 
may be able to calculate disk amplitudes involving 
closed string external states, for example. 
The results depend on whether the soliton is a D-brane or 
a ghost D-brane and we can fix which of our operators 
correspond to which.

\section*{Acknowledgements}
We would like to thank N.~Hatano,
I.~Kishimoto,
T.~Saitou, Y.~Satoh,
A.~Yamaguchi and K.~Yamamoto for discussions
and comments.
This work is supported in part
by Grants-in-Aid for Scientific Research 13135224.

\appendix
\section{Conventions}\label{sec:conventions}

\subsubsection*{$OSp$ invariant string field theory}

The reflector in the $OSp$ invariant string field theory
is given by
\begin{equation}
\left\langle R(1,2)\right|
= \delta (1,2)
\; {}_{12}\!\langle 0|
 \, e^{E(1,2)}\, \frac{1}{\alpha_1}~,
 \label{eq:reflector}
\end{equation}
where
\begin{eqnarray}
{}_{12}\! \langle 0| 
&=& {}_{1}\!\langle 0| {}_{2}\!\langle 0|~,
 \nonumber\\
E(1,2)
 &=&
  -\sum_{n=1}^\infty\frac{1}{n}
        \left(\alpha_{n}^{N(1)} \alpha_{n}^{M(2)}
        +\tilde{\alpha}_{n}^{N (1)} \tilde{\alpha}_{n}^{M(2)}
        \right)\eta_{NM}~,
\nonumber\\
\delta (1,2)
&=&
2 \delta (\alpha_1+\alpha_2)
(2\pi )^{26}\delta^{26}(p_1+p_2)
i (\bar{\pi}_0^{(1)}+\bar{\pi}_0^{(2)})
(\pi_0^{(1)}+\pi_0^{(2)})~.
\label{eq:reflector1.5}
\end{eqnarray}
The BPZ conjugate $\langle \Phi |$ of $|\Phi\rangle$
is defined as 
\begin{equation}
{}_{2} \! \langle \Phi |
=
\int d1\, \langle R(1,2)|\Phi \rangle_1~.
\label{eq:BPZ}
\end{equation}

%
%

The $\ast$-product of the string fields is defined by using 
\begin{eqnarray}
\left\langle V_3(1,2,3)\right|
=
\delta (1,2,3)
   \; {}_{123}\!\langle 0|e^{E (1,2,3)}
    C(\rho_I)
    \mathcal{P}_{123}
    \frac{|\mu (1,2,3)|^2}{\alpha_1\alpha_2\alpha_3}~,
\label{eq:V3}
\end{eqnarray}
where $\rho_{I}$ denotes the interaction point and
\begin{eqnarray}
{}_{123}\!\langle 0|
 &=& {}_{1}\!\langle 0 |\,{}_{2}\!\langle 0|\,{}_{3}\!\langle 0|~,
\nonumber\\
\mathcal{P}_{123}&=&\mathcal{P}_{1}\mathcal{P}_{2}\mathcal{P}_{3}~,
\quad \mathcal{P}_{r}=\int^{2\pi}_{0} \frac{d\theta}{2\pi}
      e^{i\theta \left(L^{(r)}_{0}-\tilde{L}^{(r)}_{0}\right)}~,
\nonumber\\
\delta (1,2,3)
&=&
2 \delta \left( \sum_{s=1}^3 \alpha_s \right)
 (2\pi )^{26} \delta^{26} \left( \sum_{r=1}^3 p_r \right)
\, i\left(\sum_{r^\prime =1}^3\bar{\pi}_0^{(r^\prime )}\right)
\left( \sum_{s^\prime =1}^3\pi_0^{(s^\prime )} \right)~,
\nonumber\\
E(1,2,3)
&=&
\frac{1}{2}\sum_{n,m\geq 0}\sum_{r,s}
\bar{N}_{nm}^{rs}
 \left( \alpha_n^{N (r)} \alpha_{m}^{M (s)}
        +\tilde{\alpha}_n^{N (r)} \tilde{\alpha}_{m}^{M (s)}
  \right)\eta_{NM}~,
\nonumber \\
\mu(1,2,3) & = &
  \exp\left(-\hat{\tau}_{0} \sum_{r=1}^{3}\frac{1}{\alpha_{r}} 
      \right)~,
    \quad \hat{\tau}_{0}
    =\sum_{r=1}^{3} \alpha_{r} \ln \left|\alpha_{r}\right|~.
\label{eq:V3variables}
\end{eqnarray}
Here $\bar{N}^{rs}_{nm}$ denote the Neumann coefficients
associated with the joining-splitting type of three-string
interaction \cite{Kaku:1974zz}%
\cite{Mandelstam:1973jk}\cite{Cremmer:1974ej}.
Notice that $\left\langle V_3(1,2,3)\right|$ is not equal to the 
three-string vertex eq.(\ref{V30}). 
The $\ast$-product $\Phi \ast \Psi$
of two arbitrary closed string fields
$\Phi$ and $\Psi$ is given as
\begin{equation}
\left|\Phi *\Psi \right\rangle_4
=
\int d1d2d3\, \left\langle V_3(1,2,3)\left|
     \Phi\right\rangle_1 \right.
     \left|\Psi \right\rangle_2
     \left| R(3,4) \right\rangle~.
 \label{eq:star-prod}
\end{equation}

\subsubsection*{Light-cone gauge string field theory}
Various quantities appearing 
in the light-cone gauge string field theory action 
can be defined in a quite similar way. 
The reflector $^{\mathrm{LC}}\left\langle R(1,2)\right|$ 
is given by
\begin{equation}
^{\mathrm{LC}}\left\langle R(1,2)\right|
= \delta_{\mathrm{LC}} (1,2)
\; {}_{12}\!\langle 0|
 \, e^{E_{\mathrm{LC}}(1,2)}\, \frac{1}{\alpha_1}~,
 \label{eq:reflectorLC}
\end{equation}
where
\begin{eqnarray}
{}_{12}\! \langle 0| 
&=& {}_{1}\!\langle 0| {}_{2}\!\langle 0|~,
 \nonumber\\
E_{\mathrm{LC}}(1,2)
 &=&
  -\sum_{n=1}^\infty\frac{1}{n}\sum_{i=1}^{24}
        \left(\alpha_{n}^{i(1)} \alpha_{n}^{i(2)}
        +\tilde{\alpha}_{n}^{i (1)} \tilde{\alpha}_{n}^{i(2)}
        \right)~,
\nonumber\\
\delta_{\mathrm{LC}} (1,2)
&=&
2 \delta (\alpha_1+\alpha_2)
(2\pi )^{25}\delta^{24}(p_1+p_2)
~.
\end{eqnarray}

The three-string vertex $^{\mathrm{LC}}\left\langle V_3^0(1,2,3)\right|$ 
can be defined as
\begin{eqnarray}
^{\mathrm{LC}}\left\langle V_3^0(1,2,3)\right|
=
\delta_{\mathrm{LC}} (1,2,3)
   \; {}_{123}\!\langle 0|e^{E_{\mathrm{LC}} (1,2,3)}
    \mathcal{P}^{\mathrm{LC}}_{123}
    \frac{|\mu (1,2,3)|^2}{\alpha_1\alpha_2\alpha_3}~,
\label{eq:V3LC}
\end{eqnarray}
where 
\begin{eqnarray}
{}_{123}\!\langle 0|
 &=& {}_{1}\!\langle 0 |\,{}_{2}\!\langle 0|\,{}_{3}\!\langle 0|~,
\nonumber\\
\mathcal{P}^{\mathrm{LC}}_{123}
 &=&\mathcal{P}^{\mathrm{LC}}_{1}
    \mathcal{P}^{\mathrm{LC}}_{2}
    \mathcal{P}^{\mathrm{LC}}_{3}~,
\quad \mathcal{P}^{\mathrm{LC}}_{r}
      =\int^{2\pi}_{0} \frac{d\theta}{2\pi}
      e^{i\theta \left(L^{\mathrm{LC}(r)}_{0}
                 -\tilde{L}^{\mathrm{LC}(r)}_{0}\right)}~,
\nonumber\\
\delta_{\mathrm{LC}} (1,2,3)
&=&
2 \delta \left( \sum_{s=1}^3 \alpha_s \right)
 (2\pi )^{25} \delta^{24} \left( \sum_{r=1}^3 p_r \right)
~,
\nonumber\\
E_{\mathrm{LC}}(1,2,3)
&=&
\frac{1}{2}\sum_{n,m\geq 0}\sum_{r,s}\sum_{i=1}^{24}
\bar{N}_{nm}^{rs}
 \left( \alpha_n^{i (r)} \alpha_{m}^{i (s)}
        +\tilde{\alpha}_n^{i (r)} \tilde{\alpha}_{m}^{i (s)}
  \right)~,
\nonumber \\
\mu(1,2,3) & = &
  \exp\left(-\hat{\tau}_{0} \sum_{r=1}^{3}\frac{1}{\alpha_{r}} 
      \right)~,
    \quad \hat{\tau}_{0}
    =\sum_{r=1}^{3} \alpha_{r} \ln \left|\alpha_{r}\right|~.
\end{eqnarray}


\end{document}